\documentclass[conference]{IEEEtran}
\IEEEoverridecommandlockouts
\usepackage{cite}
\usepackage{amsmath,amssymb,amsfonts}
\usepackage{algorithmic}
\usepackage{graphicx}
\usepackage{textcomp}
\usepackage{xcolor}
\usepackage{hyperref}
\usepackage{pifont}
\usepackage{color}
\usepackage{booktabs} 
\def\BibTeX{{\rm B\kern-.05em{\sc i\kern-.025em b}\kern-.08em
    T\kern-.1667em\lower.7ex\hbox{E}\kern-.125emX}}

\begin{document}

\title{FDG-Diff: Frequency-Domain-Guided Diffusion Framework for Compressed Hazy Image Restoration}

\author{\textbf{Ruicheng Zhang}$^1$, \textbf{Kanghui Tian}$^1$, \textbf{Zeyu Zhang}$^2$, \textbf{Qixiang Liu}$^1$, \textbf{Zhi Jin}$^{1*}$\\$^1$Sun Yat-sen University $^2$The Australian National University\thanks{$^{*}$Corresponding author: jinzh26@mail.sysu.edu.cn.}}

\maketitle

\begin{abstract}
In this study, we reveal that the interaction between haze degradation and JPEG compression introduces complex joint loss effects, which significantly complicate image restoration. Existing dehazing models often neglect compression effects, which limits their effectiveness in practical applications. To address these challenges, we introduce three key contributions. First, we design FDG-Diff, a novel frequency-domain-guided dehazing framework that improves JPEG image restoration by leveraging frequency-domain information. Second, we introduce the High-Frequency Compensation Module (HFCM), which enhances spatial-domain detail restoration by incorporating frequency-domain augmentation techniques into a diffusion-based restoration framework. Lastly, the introduction of the Degradation-Aware Denoising Timestep Predictor (DADTP) module further enhances restoration quality by enabling adaptive region-specific restoration, effectively addressing regional degradation inconsistencies in compressed hazy images. Experimental results across multiple compressed dehazing datasets demonstrate that our method consistently outperforms the latest state-of-the-art approaches.
Code be available at \url{https://github.com/SYSUzrc/FDG-Diff}.
\end{abstract}

\begin{IEEEkeywords}
Image dehazing, JPEG compression, Frequency-domain augmentation, Diffusion model
\end{IEEEkeywords}

\vspace{-0.2cm}
\section{Introduction}
\label{sec:intro}

Urban traffic cameras are often located in haze-prone areas like urban arterials and transportation hubs, capturing real-time traffic conditions for remote monitoring and analysis. To improve visual quality, efforts have focused on restoring hazy images. However, due to bandwidth and storage constraints, images are typically compressed before transmission, causing irreversible information loss and visual degradation. This degradation amplifies haze-related impacts, resulting in a complex joint loss that challenges effective restoration. Previous dehazing studies \cite{2011Single,Wang_Yu,dehazeformer,9879191,FCB,FFA,yu2024highqualityimagedehazingdiffusion} have largely neglected compression effects, limiting their applicability in real-world scenarios. Therefore, expanding image dehazing research to address the restoration of compressed hazy images is crucial. As the most widely used image compression format, JPEG is chosen in this paper for further discussion and research.

\begin{figure}[htbp]
    \centering
    \vspace{-0.25cm}
    \includegraphics[width=0.5\textwidth]{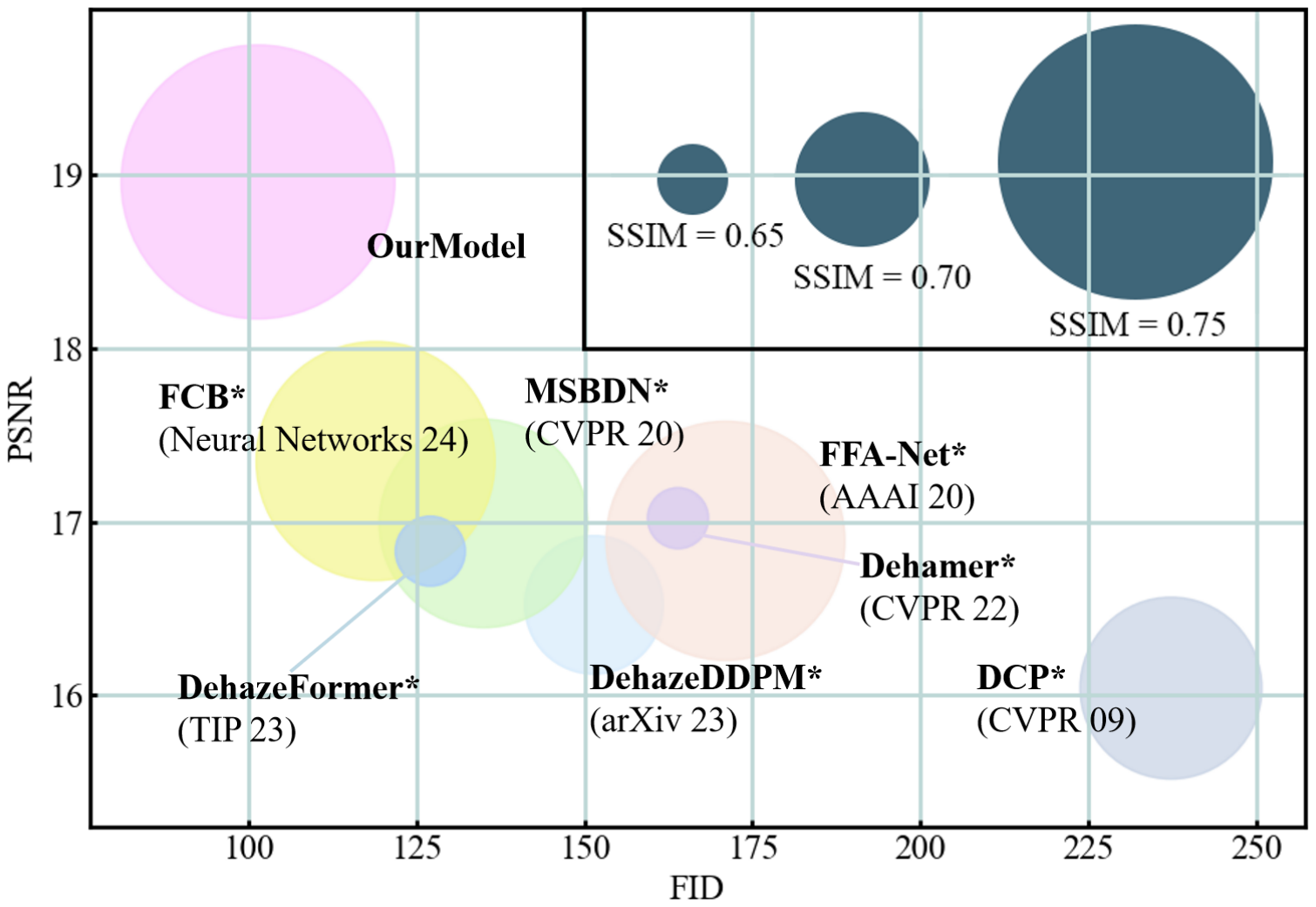}
    \setlength{\abovecaptionskip}{-0.55cm}   
    \setlength{\belowcaptionskip}{-0.6cm}   
    \caption{Improvement of FDG-Diff over the SOTA approaches on I-Haze \cite{ihaze} dataset. The circle size represents the corresponding SSIM \cite{psnrssim} values. Higher PSNR and SSIM indicate better performance, while lower FID values are preferred.}
    \label{fig:img1}
    \vspace{-0.6cm}
\end{figure}

Many dehazing methods \cite{dehazeformer,Wang_Yu,FCB,9879191,FFA,yu2024highqualityimagedehazingdiffusion} have achieved remarkable performance on widely used datasets primarily consisting of uncompressed images. However, the interaction between JPEG compression and haze-induced degradation introduces complex joint loss effects that significantly complicate restoration. This challenge stems from the JPEG quantization process, which attenuates a substantial portion of high-frequency signals to zero. Given that haze naturally reduces contrast and blurs fine details, the high-frequency components of hazy images are already weakened, making them particularly vulnerable to the adverse effects of JPEG compression. In extreme cases, this results in the near-complete loss of crucial high-frequency details that are essential for effective restoration. Existing methods largely overlook this factor, hindering their efficacy for compressed hazy images.

Denoising Diffusion Probabilistic Models (DDPMs) \cite{DDPMs} have shown remarkable success in learning complex image distributions and generating realistic details, making them promising for addressing the joint loss effects caused by JPEG compression and haze degradation. However, deep neural networks such as U-Net exhibit a spectral bias \cite{zhang2025motion}, inherently favoring low-frequency function learning during training, which complicates and slows the learning of high-frequency modes \cite{bias_DNN}. This limitation  is particularly pronounced in compressed hazy images, where high-frequency information is already scarce. Enhancing dehazing diffusion models to effectively reconstruct high-frequency signals is thus crucial.

In this work, we propose FDG-Diff, a frequency-domain-guided patch-based dehazing diffusion model with compression awareness to address the challenges. Our approach integrates frequency-domain augmentation techniques within a diffusion-based restoration framework, effectively reconstructing images and compensating for high-frequency losses. FDG-Diff consists of a spectrum decomposition network and a compression-aware frequency compensation DDPM. The spectrum decomposition network initially separates compression loss effects in the frequency domain and produces a corrected hazy image. The corrected hazy image and the compression effect spectrum serve as the conditional input for the DDPM and attention guidance for the High-Frequency Compensation Module (HFCM), respectively. Embedded within the DDPM’s cross-layer connections, the HFCM utilizes wavelet sampling to extract high-frequency features from the feature maps and performs cross-attention \cite{cross_attention} with the compression effect spectrum. This design  compensates for high-frequency losses caused by joint degradation, significantly enhancing fine image detail restoration.

Furthermore, compression loss exhibits regional variability within images, with dense haze areas suffering more significant signal loss during quantization (see in section \ref{analysis}). This interaction intensifies regional degradation inconsistencies, further complicating the restoration process. To address this issue, we introduce a region-customized restoration strategy featuring a Degradation-Aware Denoising Timestep Predictor (DADTP) module. The DADTP predicts denoising timestep offsets for each patch by leveraging transmission maps that encode regional degradation levels. By dynamically adjusting denoising intensity and timesteps across patches, the DADTP effectively mitigates the challenges posed by regionally inconsistent degradation in compressed hazy images.

Our contributions are summarized as follows:
\begin{itemize}
 \item To our knowledge, we are the first to explore dehazing on compressed images using a learning-based approach. Extensive validation reveals that JPEG compression poses substantial challenges for existing dehazing methods, highlighting the need for more research attention.

\item We propose FDG-Diff, a novel frequency-domain-guided patch-based dehazing diffusion framework for compressed hazy images restoration. FDG-Diff first separates compression effects and lossless information distributions in the frequency domain and then uses them to guide DDPM sampling. Experimental results show a substantial improvement over existing dehazing methods on JPEG images.

\item We design a High-Frequency Compensation Module (HFCM) that incorporates a cross-attention mechanism to utilize frequency-domain features for enhancing spatial-domain detail restoration. This enables precise high-frequency compensation with compression awareness.

\item To address regional inconsistencies in the degradation of compressed hazy images, we propose a patch-based diffusion restoration framework. It incorporates a Degradation-Aware Denoising Timestep Predictor (DADTP), which adaptively adjusts denoising intensity and timesteps for each patch, enabling region-specific restoration.
\end{itemize}

\section{RELATED WORKS}
\vspace{-0.05cm}

\subsection{Single Image Dehazing}
Existing dehazing algorithms can be broadly categorized into prior-based and learning-based methods. \textbf{Prior-based methods} rely on the atmospheric scattering model (ASM) \cite{asm} and handcrafted priors \cite{2011Single}, utilizing statistical analyses of hazy and clear images to extract enhancement knowledge. However, their robustness is constrained by specific model assumptions. \textbf{Learning-based methods} use deep neural networks to predict ASM coefficient \cite{dehazeNet} or directly model the hazy-to-clear translation \cite{dehazeformer}. Recent advancements in this domain focus on designing better architectures \cite{FFA} or improving optimization \cite{FCB} to enhance dehazing performance.

\begin{figure}[tbp]
    \centering
    \vspace{-0.35cm}
    \includegraphics[width=0.5\textwidth]{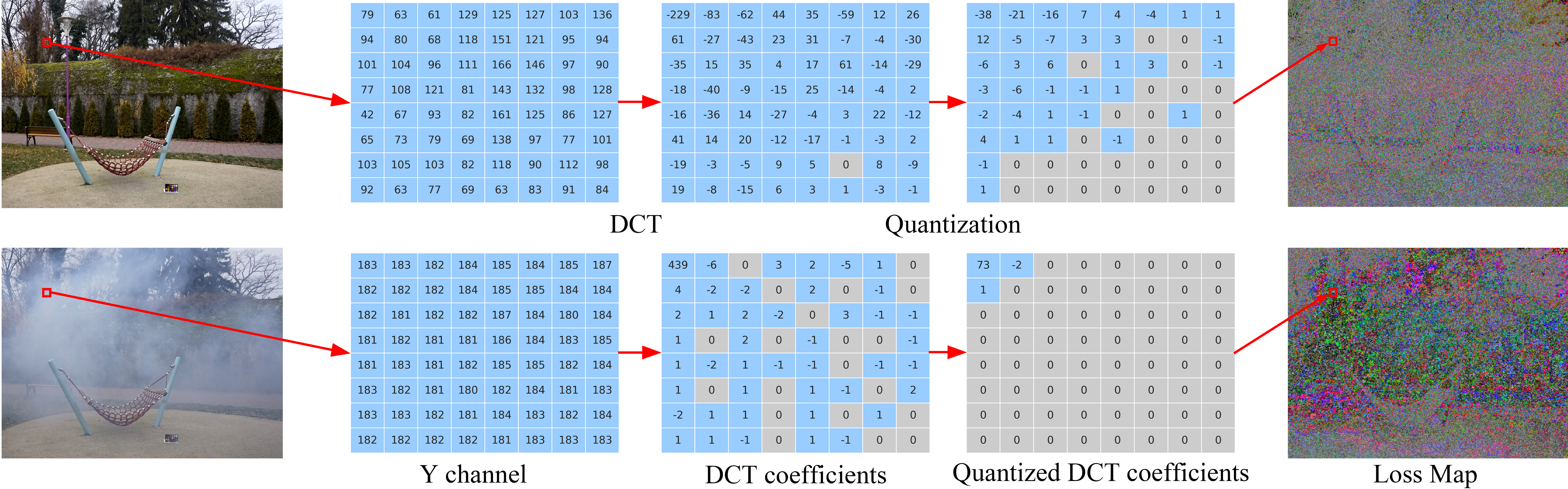}
    \setlength{\abovecaptionskip}{-0.55cm}   

\setlength{\belowcaptionskip}{-0.6cm}   
    \caption{The track of the JPEG process. The colors in the loss map reflect the magnitude of the loss, as the absolute loss is computed separately for each of the three channels.  Though compressed at a quite high QF of 80, the hazy images still suffers severe information loss in hazy regions.}
    \label{fig:long_image}
    \vspace{-0.5cm}
\end{figure}

Notably, existing learning-based approaches neglect compression effects, rendering them ineffective for compressed images. While few previous attempts \cite{investigation} have attempted to address JPEG compressed hazy images using traditional algorithms, their performance remains inferior to that of deep learning methods that fail to explicitly account for compression effects.

\vspace{-0.2cm}
\subsection{JPEG Compression}
JPEG is the most widely used format for image storage and transmission, balancing storage efficiency with quality preservation. Its core design involves discarding redundant high-frequency components in the DCT domain, which reduces storage requirements but may introduce noticeable artifacts.

Recent efforts in mitigating JPEG artifacts have achieved remarkable progress. Some approaches leverage knowledge of the compression process, such as Quality Factor  \cite{JPEG_DDPM} or quantization matrices  \cite{FBCNN}. Others focus on achieving blind restoration without JPEG-specific information \cite{DriftRec}. However, compared to other types of scenes, the inherent attenuation of high-frequency signals in hazy images makes them particularly vulnerable to severe loss during quantization and rounding operations. In extreme cases, these high-frequency components may be entirely eliminated in certain regions, presenting a critical challenge for effective image restoration.

\vspace{-0.2cm}
\subsection{Denoising Diffusion Probabilistic Models}
DDPMs have demonstrated exceptional capability in generating high-quality images both unconditionally \cite{DDPMs} and conditionally \cite{DriftRec}. Their success in image dehazing \cite{FCB} and JPEG artifact removal \cite{DriftRec} is a key source of inspiration for this work. However, to the best of our knowledge, little to no research has explored the application of diffusion models to complex restoration tasks involving joint degradation effects.

\section{PROBLEM ANALYSIS}
\vspace{-0.1cm}
\label{analysis}

In JPEG compression, the input hazy image $I\{r,g,b\} (m,n) $ is initially converted into a YUV color space $I\{y,u,v\} (m,n) $ through a linear transformation. For simplification, the analysis focuses on the primary $Y$ channel. Based on the ASM \cite{asm}, the hazy luminance can be expressed as

\vspace{-0.5em}
\begin{small}
\begin{equation}
    \hat{y}(m,n)= y(m,n)t(m,n) + a_y (1 - t(m,n)),
    \label{eq:y_hat}
    \end{equation}
\end{small}
\noindent where $y(m,n)$ represents the luminance of the non-hazy image, \( t(m,n) \) is the transmission coefficient, and $a_y$ denotes the airlight component projected onto the luminance channel.

The next step involves applying the 2D Discrete Cosine Transform (DCT) to the image. For any $N\times N$ block of the Y channel, the transformation is given by

\begin{small}
\vspace{-0.4cm}
\begin{multline}
    \hat{f}(u,v) = \alpha(u)\alpha(v) \sum_{0 \leq m, n \leq N-1} \left[ y(m,n)t(m,n)  \right. \\
    + \left. a_y(1-t(m,n)) \right] \times C_{u}(m) C_{v}(n),
    \label{eq:hazy1}
\end{multline}
\end{small}
\noindent for u, v = 0, \dots, N-1 and $ C_{s}(k) = \cos\left[\frac{(2k+1)s\pi}{2N}\right] $  and $\alpha(s) = 
\sqrt{\frac{1}{N}},  \text{ if } s = 0; \, \sqrt{\frac{2}{N}},  \text{ otherwise.} $

Assuming uniform depth across every pixel within a block $(t(m,n)=t)$, Equation (\ref{eq:hazy1}) simplifies to

\begin{small}
\vspace{-0.4cm}
\begin{multline}
\hat{f}(u,v) = 
\alpha(u)\alpha(v) \sum_{0 \leq m, n \leq N-1} \left[ y(m, n)t \right. \\
+ \left. a_y(1-t) \right] C_{u}(m) C_{v}(n),
\label{eq:hazy2}
\end{multline}
\end{small}

Next, we associate the zigzag frequency $\nu$ with the horizontal and vertical frequencies $u$ and $v$, respectively. The operators $Z_u(\nu)$, $Z_v(\nu)$, and $Z_{\nu}(u,v)$ are defined as follows:

\begin{small}
\begin{equation}
    u=Z_u(\nu),\quad v=Z_v(\nu),\quad\nu=Z_\nu(u,v),
\end{equation} 
\end{small}

 This allows us to parameterize Equation (\ref{eq:hazy2}) with $\nu$:
\begin{small}
\begin{equation}
    \hat{f}^{z}(\nu)=\hat{f}\left(Z_{u}(\nu),Z_{v}(\nu)\right).
\end{equation}
\end{small}

 Ignoring the direct-current (DC) component \( \hat{f}^z(0) \) and, hence, the effect of the addition of \( (1 - t) \alpha_y \), the alternating-current (AC) components with \( 1 \leq \nu \leq N^2 - 1 \) are
 
 \begin{small}
 \vspace{-0.4cm}
\begin{equation}
    \begin{aligned}\hat{f}^{z}(\nu)=&\alpha(Z_{\nu}(u))\alpha(Z_{\nu}(v))\sum_{0\leq m,n\leq N-1}y(m,n)t\\&\times C_{u}(\nu_{m})C_{v}(\nu_{n}) =tf^{z}(\nu), \hspace{0.1cm} \text{for } 1\leq\nu\leq N^{2}-1.\end{aligned}
    \label{eq:hazy3}
\end{equation}
\end{small}
As shown in (\ref{eq:hazy3}), the AC components of the hazy block \( \hat{f}^z\)  are attenuated by the transmission coefficient \(t\) compared to their non-hazy counterpart \( f^z\). When quantization is applied to each DCT coefficient, we have

\begin{small}
\begin{equation}
\hat{f}^q(\nu)=\left\lfloor\hat{f}^z(\nu)/q(\nu)+1/2\right\rfloor 
    \label{eq:lianghua},
\end{equation} 
\end{small}
\noindent where operator $\lfloor x+1/2 \rfloor$ rounds $x$ to the nearest integer and $q(\nu)$ represents the quantization matrix. After the quantization process  in Equation \ref{eq:lianghua}, an AC component is annihilated when $ h^z(\nu)/q(\nu)<1/2 $.

The probability of a non-hazy AC coefficient being annihilated can be expressed as
$ P\left[\left|f^z(\nu)\right|<\frac{q(\nu)}{2}\right] $. Similarly, the the probability that a hazy AC coefficient is annihilated as

\begin{small}
\begin{equation}
    \begin{aligned}
P\left[\left|\hat{f}^{z}(\nu)\right|<\frac{q(\nu)}{2}\right]=P\left[\left|f^{z}(\nu)\right|<\frac{q(\nu)}{2t}\right].
\end{aligned}
\end{equation}
\end{small}

The values of transmission coefficient $t$ range from 0 to 1, which gives us the following inequality:

\vspace{-0.35em}
\begin{small}
\begin{equation}
    P\left[\left|\hat{f}_{i,j}^z(\nu)\right|<\frac{q(\nu)}2\right]>P\left[\left|f_{i,j}^z(\nu)\right|<\frac{q(\nu)}2\right].
    \label{eq:conclusion}
\end{equation}
\end{small}

\vspace{-0.6em}
Thus, the presence of haze increases the likelihood that the AC coefficients of the image will be reduced to zero during the JPEG compression process. This joint loss is positively correlated with the density of the haze, manifesting as increased loss of high-frequency signals and image details.

\begin{figure*}[htbp]
    \centering
    \vspace{-0.35cm}
    \setlength{\abovecaptionskip}{-0.05cm}   
    \includegraphics[width=0.9\textwidth]{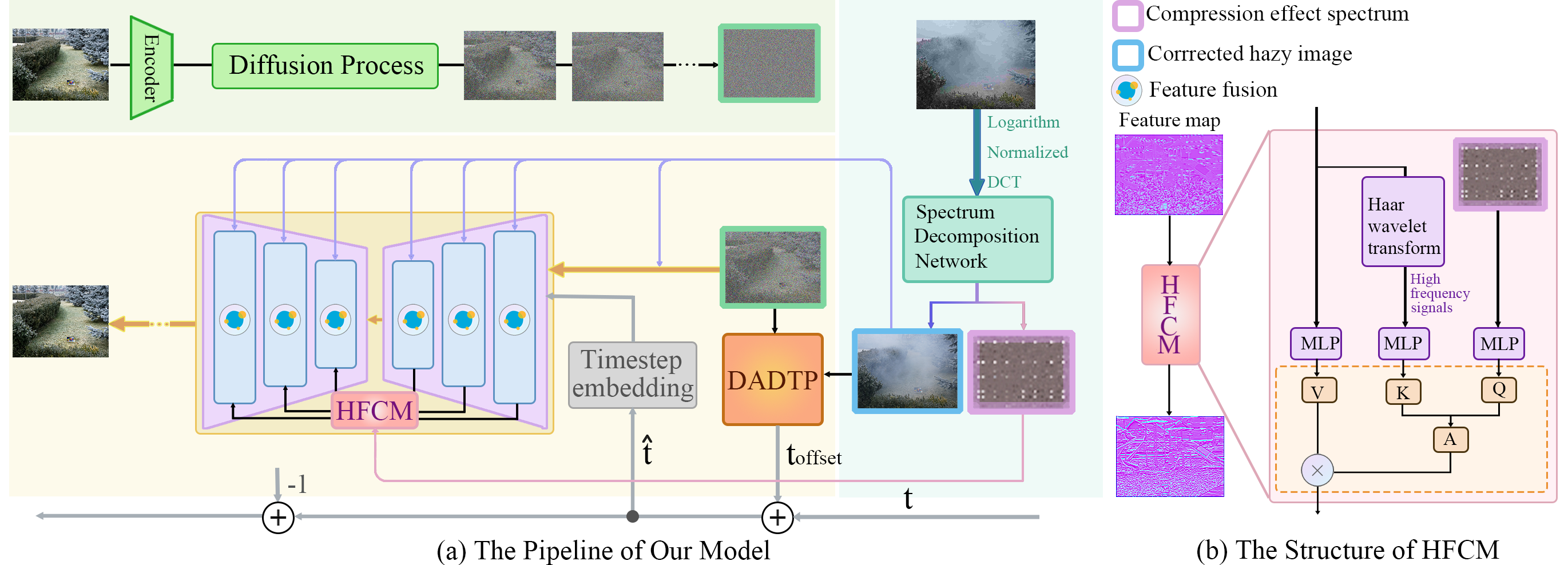}

\setlength{\belowcaptionskip}{-1.4cm}   
 \caption{(a) The pipeline of FDG-Diff, comprising a spectrum decomposition network and a compression-aware frequency compensation DDPM. The spectral decomposition network separates compression  effects in the frequency domain and generates a corrected hazy image, which serves as the conditional guidance for the DDPM. Additionally, the Degradation-Aware Denoising Timestep Predictor (DADTP) adjusts the denoising timesteps for each patch to enable region-specific restoration. (b) The detailed block design of the HFCM. Input features are enhanced in the mid-to-high frequency band using the wavelet transform and processed through cross-attention \cite{cross_attention} with the compression effect spectrum to achieve precise compensation for the signal loss caused by compression.}
    \label{fig:pipeline}
    \vspace{-0.35cm}
\end{figure*}

\vspace{-0.4em}
\section{METHOD}
\subsection{Spectrum Decomposition Network}
Our task involves two types of degradation, with one being JPEG compression, which can be easily simulated. This enables us to initially separate compression effects and lossless information, serving as the subsequent restoration guidance. Previous studies \cite{DCT_JPEG} have shown that the DCT coefficients, as the native representation of JPEG compression, offer a direct and efficient means to model compression effects. So, the spectrum decomposition network operates in the DCT domain.

Mathematically, the compression effect can be  represented as a singular matrix $\phi$, with the compressed hazy image $I^{c}$ expressed as

\begin{small}
\vspace{-0.5cm}
\begin{equation}
I^{c} = \phi^{T}I, 
\vspace{-0.2cm}
\end{equation} 
\end{small}
where $I$ represents the uncompressed hazy image.

To facilitate the model's learning process, we apply a logarithmic transformation to convert the multiplicative composition into an additive form:

\begin{small}
\vspace{-0.2cm}
\begin{equation}
ln(I^{c} ) = ln(\phi^{T})+ln(I).    
\end{equation} 
\end{small}
 
 Following normalization, the Discrete Cosine Transform (DCT) is applied to individual $8\times8$ patches, consistent with the localized quantization process in JPEG compression:
 
\begin{small}
\vspace{-0.25cm}
\begin{equation}
    D_{1}\left[\operatorname{Nom}\left(\ln I^{c}\right)\right]=D_{2}[\operatorname{Nom}(\ln \phi^{T})]+D_{3}[\operatorname{Nom}(\ln I)],
\end{equation}  
\end{small}
\noindent where \( D[\cdot] \) represents the DCT operation, and \( \text{Nom}(\cdot) \) denotes the normalization operation. 

The decomposition network is built upon the EfficientNet backbone \cite{EfficientNet} and is designed to learn the compression effect \( D[\text{Nom}(\ln \phi^{T})] \) and the corrected hazy image \( D[\text{Nom}(\ln I)] \) from the compressed hazy image  \( D[\text{Nom}(\ln I^c)] \) in the DCT domain. To optimize the network, we employ the Charbonnier loss function, defined as

\begin{small}
\vspace{-0.35cm}
\begin{equation}
\mathcal{L} = \sqrt{(D_{2}^{\text{pred}} - D_{2}^{\text{GT}})^2 + \epsilon^2} + \sqrt{(D_{3}^{\text{pred}} - D_{3}^{\text{GT}})^2 + \epsilon^2},
\end{equation}
\end{small}
\noindent where \( D_{2}^{\text{GT}} \) and \( D_{3}^{\text{GT}} \) represent the ground truth values of the compression effect and the uncompressed hazy image in the DCT domain, respectively. These ground truth values are easily derived by compressing images from a hazy image dataset. The constant \( \epsilon \) is set to \( 10^{-3} \).

\subsection{Compression-Aware Frequency Compensation DDPM}
Unlike other forms of image degradation, the joint loss  caused by compression and haze degradation results in a substantial reduction of high-frequency information. However, deep neural networks (DNNs),, such as U-Net, exhibit a spectral bias favoring low-frequency signals \cite{bias_DNN}, prioritizing the learning of smooth, low-frequency modes while delaying the recovery of high-frequency components. This limitation is particularly pronounced in compressed hazy images, where high-frequency information guidance is already scarce, further complicating the restoration of image details. Wang et al. \cite{FCB} integrate high-pass filters into U-Net skip connections to promote the transmission of high-frequency signals. However, such enhancements lack selectivity, indiscriminately amplifying both valid high-frequency details and noise. In contrast, enabling DDPM to explicitly perceive compression-induced losses and perform targeted high-frequency enhancements presents a more effective solution.

To this end, we propose a novel compression-aware frequency compensation DDPM framework, which synergistically integrates with the spectrum decomposition network. This DDPM incorporates a High-Frequency Compensation Module (HFCM) that utilizes a compression-awareness cross-attention mechanism \cite{cross_attention} for perceive high-frequency modes compensation. Specifically, the HFCM employs the Haar wavelet transform at the skip connections to extract the high-frequency components $X_H$ from the feature maps $X$. These components are then processed through a cross-attention mechanism with the compression effect spectrum $X_D$ predicted by the spectrum decomposition network:

\vspace{-0.5em}
\begin{small}
\begin{equation}
\setlength\abovedisplayskip{33pt}
\setlength\belowdisplayskip{33pt}
\begin{aligned}
\mathbf{q} &= \mathbf{x}^T \mathbf{W}_q, \quad \mathbf{k} = \mathbf{x}_{\text{H}}^T \mathbf{W}_k, \quad \mathbf{v} = \mathbf{x}_{\text{D}}^T \mathbf{W}_v, \\
\mathbf{A} &= \text{softmax}(\mathbf{q} \mathbf{k}^T / \sqrt{C/h}), \quad \text{CA}(\mathbf{x}) = \mathbf{A} \mathbf{v},
\end{aligned}
\end{equation}
\end{small}

\noindent where \(\mathbf{W}_q, \mathbf{W}_k, \mathbf{W}_v \in \mathbb{R}^{C \times (C/h)}\) are learnable parameters, \(C\) and \(h\) denote the embedding dimension and the number of heads, respectively. 

In the forward process, starting with a clean image $J_0$, Gaussian noise is iteratively added over $T$ steps. The noisy version of the image at step $t$, denoted as $J_t$, is obtained by sampling noise $\epsilon \sim N(0, 1)$ and applying the following transformation:

\begin{small}
\vspace{-0.5cm}
\begin{equation}
\setlength\abovedisplayskip{6pt}
\setlength\belowdisplayskip{5pt}
J_t = \sqrt{\gamma_t} J_0 + \sqrt{1 - \gamma_t} \epsilon,
\end{equation}
\end{small}

\noindent where $\gamma_{t} = \prod_{i=1}^{t} \alpha_{i}$, and$\left\{ \alpha_{t} \in \left( 0, 1 \right) \right\}_{t=1}^{T}$ is a sequence to control the noise scale at each step $t$. 

In the reverse process, as illustrated in Figure~\ref{fig:pipeline}(a), at each timestep, our model first obtains the adjusted timestep $\hat{t}$ from the DADTP and then restores the denoised intermediate state $J_{\hat{t}-1}$ from $J_{\hat{t}}$, based on the estimated noise $\bar{\epsilon}_{\hat{t}}$, using the following update rule:

\begin{small}
\vspace{-0.2cm}
\begin{equation}
\setlength\abovedisplayskip{0pt}
\setlength\belowdisplayskip{2pt}
 J_{\hat{t}-1}=\frac{1}{\sqrt{\alpha_{\hat{t}}}}\left(J_{\hat{t}}-\frac{1-\alpha_{\hat{t}}}{\sqrt{1-\gamma_{\hat{t}}}} f_{\theta}\left(\widetilde{I}, J_{\hat{t}}, \gamma_{\hat{t}}\right)\right)+\sqrt{1-\alpha_{\hat{t}}} \bar{\epsilon}_{\hat{t}}.
\end{equation}
\end{small}

Training the reverse process involves training the noise estimation U-Net $f_{\theta} $ to predict $\epsilon_{\hat{t}} $ given $J_{\hat{t}}$ and $\gamma_{\hat{t}}$ with the corrected hazy image $\widetilde{I}$ as an additional input. The loss function, based on the $L_{1}$ norm, is defined as

\begin{small}
\vspace{-0.3cm}
\begin{equation}
\setlength\abovedisplayskip{6pt}
\setlength\belowdisplayskip{0pt}
\mathbb{E}_{(I, J)} \mathbb{E}_{\epsilon \sim \mathcal{N}(0,1), \gamma_{l}}\|f_{\theta}(I, \underbrace{\sqrt{\gamma_{\hat{t}}} J_{0}+\sqrt{1-\gamma_{\hat{t}}} \epsilon_{\hat{t}}}_{J_{\hat{t}}}, \gamma_{\hat{t}})-\epsilon_{\hat{t}}\|_{1}^{1}.
\end{equation}
\end{small}

\begin{figure}[htbp]
    \centering
    \vspace{-0.35cm}
    \setlength{\abovecaptionskip}{-0.05cm}   
    \includegraphics[width=0.5\textwidth]{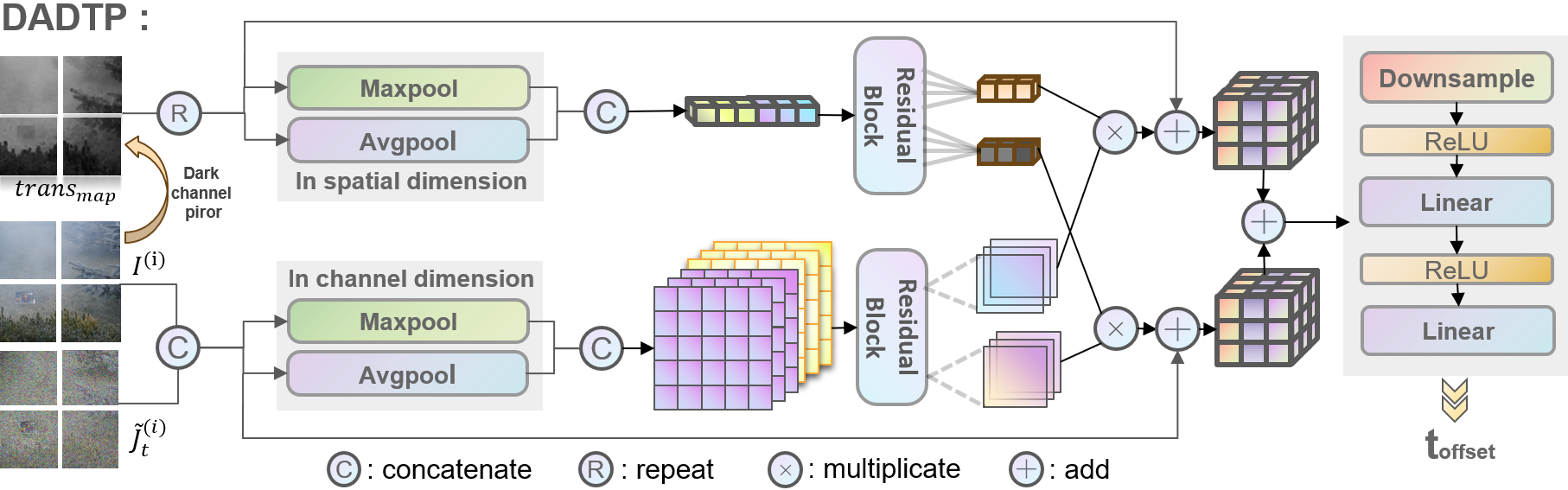}
    
\setlength{\belowcaptionskip}{-1.2cm}   
    \caption{The structure of DADTAP. The DADTP employs a spatial-channel dual-attention mechanism to effectively fuse features, adaptively adjusting the timestep offset for each patch.}
    \label{fig:long_image}
\end{figure}

\vspace{-0.20cm}
\subsection{Degradation-Aware Denoising Timestep Predictor}
Real-world images often exhibit uneven haze distribution, with haze density increasing with depth, leading to regional variations in degradation. JPEG compression further amplifies these inconsistencies, as its loss is positively correlated with fog density. Previous studies \cite{dehazeformer,Wang_Yu,FCB,9879191,FFA,yu2024highqualityimagedehazingdiffusion} generally treat all regions equally, yielding satisfactory overall results but frequently failing to reconstruct severely degraded areas effectively.

\begin{figure*}[htbp]
    \centering
    \vspace{-0.35cm}
    \setlength{\abovecaptionskip}{-0.05cm}   
    \includegraphics[width=\textwidth]{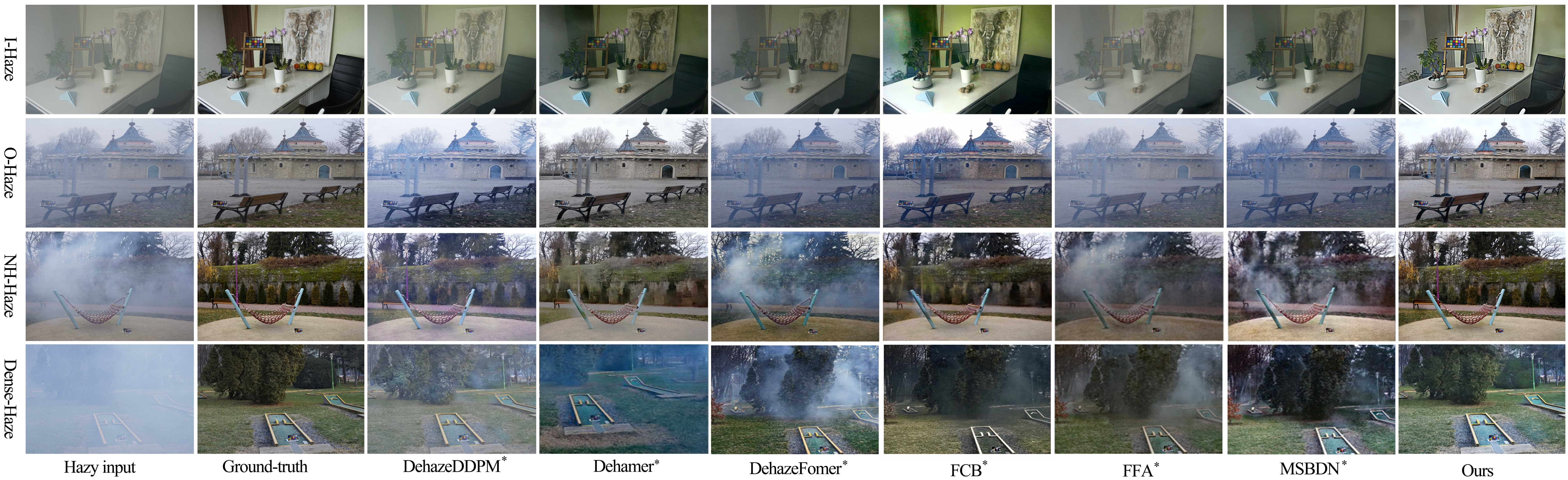}
    
\setlength{\belowcaptionskip}{-1.2cm}   
    \caption{Visual comparison on I-Haze-JPEG \cite{ihaze}, O-Haze-JPEG \cite{ohaze}, Dense-Haze-JPEG \cite{densehaze}, and NH-Haze-JPEG \cite{nhaze}, with all inputs compressed at QF 80. *Denotes the adoption of the superior cascading strategy. Our method yields fewer artifacts, more realistic details and better color consistency than others.}
    \label{fig:long_image}
    \vspace{-0.15cm}
\end{figure*}

\begin{table*}[]
\vspace{-1em}
\centering
\setlength{\tabcolsep}{2pt} 
\renewcommand{\arraystretch}{0.8} 
\caption{Experiments on I-Haze \cite{ihaze}, O-Haze \cite{ohaze}, Dense-Haze \cite{densehaze}, and NH-Haze \cite{nhaze} datasets. ↑/↓ denotes that larger/smaller values lead to better quality. Optimal and suboptimal metric values are \textbf{bolded} and \underline{underlined}, respectively. }
\label{tab:t1} 
\vspace{-1.0em}
\scalebox{0.77}{
\begin{tabular}{c|cccc|cccc|cccc|cccc}  
Datasets   & \multicolumn{4}{c|}{I-HAZE} & \multicolumn{4}{c|}{O-HAZE} & \multicolumn{4}{c|}{Dense-HAZE} & \multicolumn{4}{c}{NH-HAZE}  \\ 
 Indicators & PSNR $\uparrow$  & SSIM $\uparrow$  & LPIPS $\downarrow$ & FID $\downarrow$ & PSNR $\uparrow$  & SSIM $\uparrow$  & LPIPS $\downarrow$ & FID $\downarrow$  & PSNR $\uparrow$ & SSIM $\uparrow$ & LPIPS $\downarrow$ & FID $\downarrow$  & PSNR $\uparrow$ & SSIM $\uparrow$ & LPIPS $\downarrow$ & FID $\downarrow$ \\ 
\midrule
DCP \cite{2011Single}        & 14.58 & 0.65 & 0.55 & 310.07 & 11.64 & 0.58 & 0.62 & 280.97 & 9.57 & 0.39 & 0.87 & 409.87 & 10.51 & 0.45 & 0.66 & 192.62 \\
DehazeDDPM \cite{yu2024highqualityimagedehazingdiffusion} &   14.64 & 0.61 & 0.37 & 207.23 & 14.31 & 0.53 & 0.40 & 203.52 & 15.15 & 0.55 & 0.57 & 236.86 & 19.20 & 0.67 & 0.38 & 181.68 \\
FFA-Net \cite{FFA}   &   15.24 & 0.69 & 0.41 & 228.62 & 14.13 & 0.46 & 0.48 & 237.61 & 8.96 & 0.37 & 0.74 & 465.51 & 9.87 & 0.43 & 0.51 & 444.97 \\
MSBDN \cite{Wang_Yu}     &   15.69 & 0.72 & 0.32 & 191.78 & 13.71 & 0.56 & 0.43 & 248.36 & 8.78 & 0.35 & 0.64 & 390.92 & 10.11 & 0.44 & 0.59 & 360.03 \\
Dehamer \cite{9879191}   &   16.13 & 0.62 & 0.32 & 220.07 & 15.81 & 0.54 & 0.44 & 182.45 & 14.54 & 0.54 & 0.56 & 279.22 & 18.40 & 0.62 & 0.37 & 212.24 \\
FCB \cite{FCB}       &   16.42 & 0.68 & 0.32 & 188.30 & 14.58 & 0.48 & 0.37 & 209.95 & 10.53 & 0.45 & 0.58 & 349.50 & 11.59 & 0.58 & 0.33 & 198.43 \\
DehazeFormer \cite{dehazeformer}    & 15.37 & 0.58 & 0.49 & 185.87 & 14.44 & 0.53 & 0.43 & 195.60 & 10.30 & 0.44 & 0.66 & 444.24 & 10.87 & 0.26 & 0.49 & 376.07 \\
$ \textnormal{DCP}\rightarrow \textnormal{FBCNN}$    & 15.32 & 0.68 & 0.41 & 287.17 & 12.52 & 0.61 & 0.45 & 241.57 & 9.83 & 0.40 & 0.75 & 396.21 & 10.48 & 0.49 & 0.55 & 174.59 \\
$ \textnormal{DehazeDDPM}\rightarrow \textnormal{FBCNN}$   &  14.38 & 0.64 & 0.28 & 202.09 & 15.46 & 0.56 & 0.33 & 186.97 & 15.69 & 0.56 & 0.48 & 213.93 & 20.01 & 0.70 & 0.29 & 159.51 \\
$ \textnormal{FFA-Net}\rightarrow \textnormal{FBCNN}$      &   15.83 & 0.71 & 0.41 & 222.14 & 14.88 & 0.49 & 0.39 & 212.06 & 9.61 & 0.37 & 0.70 & 466.68 & 10.76 & 0.47 & 0.49 & 414.98 \\
$ \textnormal{MSBDN}\rightarrow \textnormal{FBCNN}$      &   15.46 & 0.73 & 0.40 & 179.17 & 14.26 & 0.59 & 0.40 & 221.59 & 8.99 & 0.36 & 0.66 & 391.73 & 10.49 & 0.46 & 0.49 & 329.47 \\
$ \textnormal{Dehamer}\rightarrow \textnormal{FBCNN}$     &   16.15 & 0.62 & 0.41 & 205.80 & 16.40 & 0.57 & 0.29 & 161.41 & 14.83 & 0.54 & 0.42 & 265.12 & 19.14 & 0.65 & 0.36 & 175.43 \\
$ \textnormal{FCB}\rightarrow \textnormal{FBCNN}$         &  16.79 & 0.72 & 0.28 & 167.66 & 15.40 & 0.51 & 0.27 & 196.84 & 11.32 & 0.47 & 0.49 & 353.51 & 12.95 & 0.59 & 0.35 & 186.97 \\
$ \textnormal{DehazeFormer}\rightarrow \textnormal{FBCNN}$     &   16.36 & 0.62 & 0.29 & 167.18 & 15.65 & 0.56 & 0.35 & 176.04 & 11.27 & 0.44 & 0.52 & 428.34 & 11.11 & 0.29 & 0.37 & 356.70 \\
$ \textnormal{FBCNN}\rightarrow \textnormal{DCP}$         & 16.04 & 0.72 & 0.33 & 237.28 & 14.62 & 0.68 & 0.35 & 196.41 & 11.75 & 0.42 & 0.60 & 343.65 & 12.49 & 0.51 & 0.42 & 128.90 \\
$ \textnormal{FBCNN}\rightarrow \textnormal{DehazeDDPM} $ &  16.52 & 0.70 & 0.17 & 151.44 & 17.02 & 0.70 & \underline{0.17} & 133.44 & \underline{18.14} & \underline{0.59} & 0.32 & \underline{171.06} & \underline{21.78} & \underline{0.72} & \underline{0.17} & \underline{121.06} \\
$ \textnormal{FBCNN}\rightarrow \textnormal{FFA-Net}$   &   16.89 & \underline{0.74} & 0.22 & 171.01 & 17.04 & 0.59 & 0.26 & 171.06 & 11.07 & 0.40 & 0.51 & 413.22 & 12.33 & 0.48 & 0.33 & 374.31 \\
$ \textnormal{FBCNN}\rightarrow \textnormal{MSBDN}$    &  16.99 & 0.73 & 0.19 & 135.01 & 16.56 & 0.65 & 0.27 & 176.99 & 15.14 & 0.52 & 0.38 & 335.03 & 12.58 & 0.48 & 0.32 & 287.81 \\
$ \textnormal{FBCNN}\rightarrow \textnormal{Dehamer}$    &   17.02 & 0.64 & 0.18 & 163.94 & \underline{18.42} & \underline{0.71} & \underline{0.17} & \underline{107.37} & 16.62 & 0.56 & \underline{0.28} & 223.65 & 20.66 & 0.68 & 0.18 & 138.49 \\
$ \textnormal{FBCNN}\rightarrow \textnormal{FCB} $   &  \underline{17.35} & \underline{0.74}& \underline{0.14} & \underline{118.95} & 17.39 & 0.64 & \textbf{0.15} & 137.70 & 16.16 & 0.59 & 0.30 & 198.42 & 17.17 & 0.62 & \underline{0.17} & 133.60 \\ 
$ \textnormal{FBCNN}\rightarrow \textnormal{DehazeFormer} $    &16.83 & 0.65 & 0.19 & 127.09 & 17.30 & 0.67 & 0.23 & 127.28 & 13.09 & 0.47 & 0.38 & 378.74 & 16.70 & 0.61 & 0.23 & 213.77    \\ \midrule 
Ours       &    \textbf{18.96} & \textbf{0.75}  & \textbf{0.12}  & \textbf{101.46}  & \textbf{20.24}  & \textbf{0.72}  & \textbf{0.15} & \textbf{104.65}  & \textbf{20.43}  & \textbf{0.62}  & \textbf{0.25}  & \textbf{132.43}  & \textbf{23.38}  & \textbf{0.81}  & \textbf{0.14}  & \textbf{84.65}   \\ \midrule 

\end{tabular}
}
\vspace{-2em}
\end{table*}

To address this issue, we introduce the Degradation-Aware Denoising Timestep Predictor (DADTP) module within a patch-based restoration framework to enable region-specific reconstruction. The module leverages the dark channel prior \cite{2011Single} transmission map, which encodes haze density information, to quantify the regional degradation degree.

During each timestep of the reverse process, DADTP predicts the timestep offset for each patch by leveraging the intermediate denoised result \( J_t \) and the transmission map \( T_{map} \). The offset is added to the current timestep, determining the new timestep for the patch, with subsequent noise estimation adjusted accordingly through timestep embedding. DADTP employs a spatial-channel dual-attention mechanism \cite{TAHM} to effectively extract features. This is followed by downsampling and fully connected layers, which output the timestep offset.

To mitigate edge artifacts resulting from inconsistent restoration across patches, we adopt a sliding window approach for sampling fusion. Specifically, a patch of size \( p \times p \) is extracted from the upper-left corner of the image and slid both vertically and horizontally across the entire image with a stride of \( r \) (\( r < p \)), creating a set of overlapping patches. At each timestep \( t \), noise estimation for individual patches is performed independently. For pixels in overlapping regions, sampling updates are guided by the averaged noise estimation \( \overline {\epsilon}_\theta \) across all overlapping patches, ensuring seamless fusion:

\begin{small}
\begin{equation}
\setlength\abovedisplayskip{3pt}
\setlength\belowdisplayskip{0pt}
\overline{\epsilon} _\theta = \frac{1}{n}  {\textstyle \sum_{i=1}^{n}} \epsilon _\theta(J_t^{(i)},\tilde I^{(i)},t).
\end{equation}
\end{small}

\section{EXPERIMENTS}
\subsection{Experiments Settings}
Following the recommendation in \cite{Morley2020}, all experiments are conducted on JPEG hazy images with a QF of 80, which provides a balance between storage efficiency and visual quality, making it representative of typical real-world scenarios.

Our method is evaluated on four benchmark dehazing datasets: I-Haze \cite{ihaze}, O-Haze \cite{ohaze}, Dense-Haze \cite{densehaze}, and NH-Haze \cite{nhaze}. These datasets cover a wide variety of natural haze conditions such as uniform haze, non-uniform haze, dense haze, and thin haze. For each dataset, the last $ 10\% $ images are reserved for testing, while the remaining images are used for training. The patch size is set to $p = 64$, with a sliding stride of $r = 16$. All experiments are conducted on a system equipped with four NVIDIA GeForce RTX 3090 GPUs.

\vspace{-0.5em}
\subsection{Quantitative Evaluation}

For comparative analysis, we evaluate several representative state-of-the-art (SOTA) dehazing methods alongside their cascaded variants, which are combined with FBCNN \cite{FBCNN}, a leading model for JPEG compression artifact removal. Two cascading strategies are employed: dehazing followed by artifact removal, and artifact removal followed by dehazing. The performance of the models is assessed using two distortion-based metrics, PSNR and SSIM \cite{psnrssim}, as well as two perceptual-based metrics, LPIPS \cite{LPIPS} and FID \cite{FID}.

As showed in Table \ref{tab:t1}, our method largely outperforms other baselines across all the compressed datasets. Notably, our model demonstrates a more substantial performance advantage in dense haze \cite{densehaze} and uneven haze \cite{nhaze} scenarios. This superiority is mainly attributed to the powerful high-frequency compensation capability of the HFCM and the loss-aware, region-customized restoration enabled by the DADTP.

Additionally, the results highlight that simple cascading strategies fail to effectively address the dual degradation caused by haze and JPEG compression, underscoring the importance of investigating this challenging problem. Another interesting observation is that the cascading strategy of artifact removal followed by dehazing outperforms the reverse order. This finding further justifies our approach to first mitigate the impact of JPEG compression and then incorporate it into the iterative process of the subsequent diffusion model.

\vspace{-0.2cm}
\subsection{Qualitative Evaluation}
\vspace{-0.2cm}
We present several samples in Fig.\ref{fig:long_image} from the test set to compare our method with other SOTA approaches. Here, we only showcase the results of the better cascading strategy. Due to the impact of compression, the results of other methods exhibit severe artifacts and noise, along with noticeable color distortions. Our results demonstrate superior visual quality, with significantly fewer artifacts and noise and better color consistency, highlighting the effectiveness of our approach in mitigating the effects of compression.

\begin{table}[h]
\vspace{-1.5em}
\centering
\small 
\caption{Ablation Experiments}
\label{tab:ablation}
\resizebox{0.5\textwidth}{!}{  
\renewcommand{\arraystretch}{1.5}  
\begin{tabular}{*{6}{c}}  
    \toprule
    \large HFCM  & \large DADTP & \large PSNR & \large SSIM & \large LPIPS & \large FID   \\
    \midrule
{\normalsize \ding{55}} & {\normalsize \ding{55}}  & {\normalsize 19.05} & {\normalsize 0.74} & {\normalsize 0.21}& {\normalsize 136.67}   \\
    {\normalsize \ding{55}}  & {\normalsize \ding{51}}  & {\normalsize$20.34(\textcolor{red}{6.77\%\uparrow})$} & {\normalsize $0.76(\textcolor{red}{2.72\%\uparrow})$}& {\normalsize $0.18(\textcolor{red}{14.29\%\downarrow)}$} & {\normalsize $106.48(\textcolor{red}{22.09\%\downarrow})$}  \\
    {\normalsize \ding{51}}  & {\normalsize \ding{55}}  & {\normalsize $21.89 (\textcolor{red}{14.91\%\uparrow})$} & {\normalsize $0.77(\textcolor{red}{4.05\%\uparrow})$} & {\normalsize $0.15(\textcolor{red}{28.57\%\downarrow})$} & {\normalsize $95.32(\textcolor{red}{30.26\%\downarrow})$}  \\
    {\normalsize \ding{51}}  & {\normalsize \ding{51}}  & {\normalsize $23.38 (\textcolor{red}{22.73\%\uparrow})$} & {\normalsize $0.81(\textcolor{red}{9.46\%\uparrow})$} & {\normalsize $0.14(\textcolor{red}{33.33\%\downarrow})$} & {\normalsize $84.65(\textcolor{red}{38.06\%\downarrow})$}   \\
    \bottomrule
\end{tabular}
}
\vspace{-1em}
\end{table}

\vspace{-0.2cm}
\subsection{Ablation Study}
\vspace{-0.2cm}
We perform ablation studies on various architectural configurations of our model to investigate the contribution of each component. All tests are conducted on the NH-Haze dataset. The results (Table \ref{tab:ablation}) demonstrate that both the HFCM and the DADTP significantly improve image restoration quality with their combination achieving the optimal results.

\begin{figure}[htbp]
    \centering
    \vspace{-0.35cm}
    \setlength{\abovecaptionskip}{0.cm}
    \includegraphics[width=0.5\textwidth]{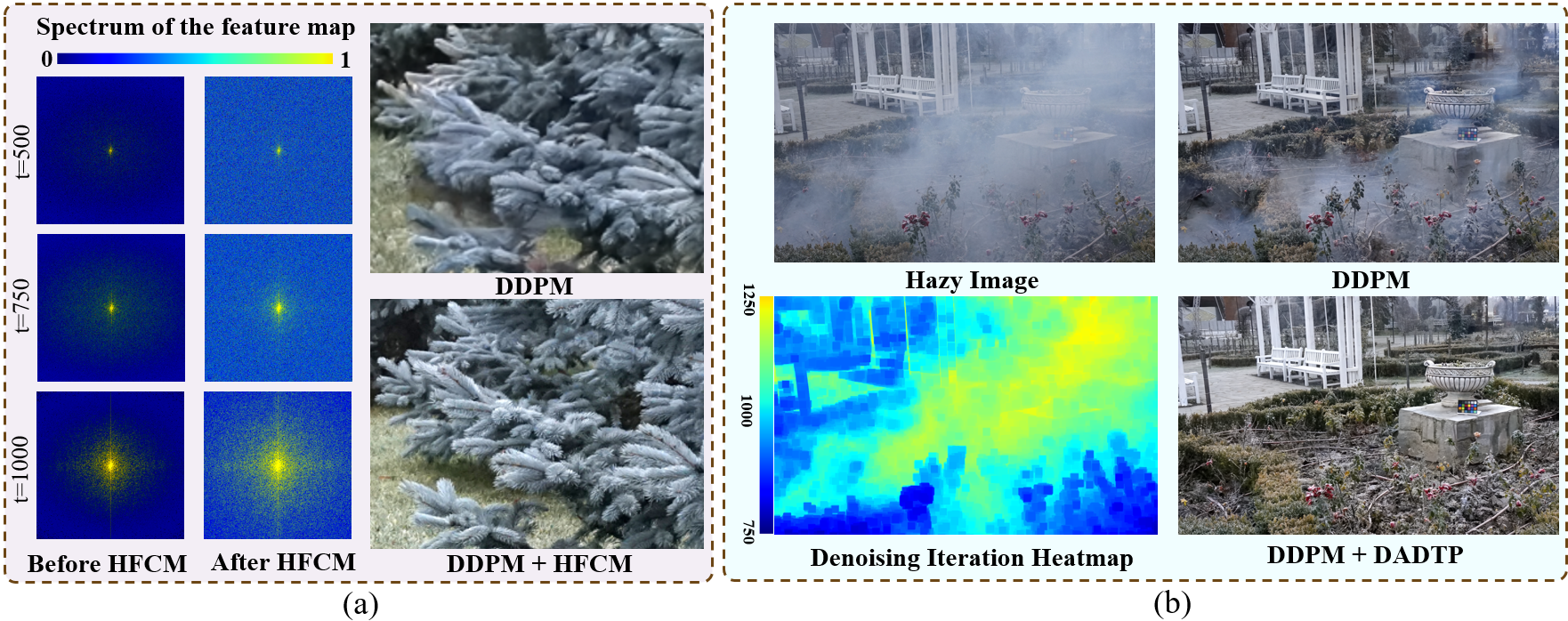}
    
    \caption{Ablation study visualizations. Figure (a) illustrates the high-frequency detail compensation capability of the HFCM through the power spectrum of feature maps before and after processing at different timesteps. Figure (b) highlights DADTP's ability to effectively restore inconsistently degraded regions by dynamically adjusting the denoising timesteps and intensity.
}
    \label{fig:ablation}
    \vspace{-0.35cm}
\end{figure}

\section{CONCLUSION}
In this paper, we identify the challenges posed by  compressed hazy images and propose a comprehensive solution. Leveraging the characteristics of compressed hazy images, we design FDG-Diff, a frequency-domain-guided patch-based dehazing diffusion model that includes a spectrum decomposition network and a compression-aware frequency compensation DDPM. We find that diffusion models  benefit from effectively frequency-domain augmentation, offering potential utility for other inverse problems. The introduction of the DADTP module further enhances restoration quality by adapting to regional variations in image degradation. Experimental evaluations on multiple compressed dehazing datasets reveal that our method consistently outperforms existing SOTA approaches. 


\end{document}